\documentclass[a4paper,10pt,twoside]{cpc-hepnp}
\usepackage{multicol}
\usepackage{graphicx}
\usepackage{booktabs}
\usepackage{amssymb,bm,mathrsfs,bbm,amscd}
\usepackage[tbtags]{amsmath}
\usepackage{lastpage}
\usepackage{CJK}

\begin{document}
\begin{CJK*}{GBK}{song}

\fancyhead[co]{\normalsize
}
\footnotetext[0]{ }

\title{\large $I=1/2$ low-lying mesons in lattice QCD
\thanks{
Supported by the National Natural Science Foundation of China (11175124) and
the National Magnetic Confinement Fusion Program of China (2013GB109000).
}
}
\author{
\quad Ziwen Fu$^{1}$\email{fuziwen@scu.edu.cn}
}
\maketitle

\address{%
$^1$ Key Laboratory of Radiation Physics and Technology {\rm (Sichuan University)}, Ministry of Education;
Institute of Nuclear Science and Technology, College of Physical Science and Technology,
Sichuan University, Chengdu 610064, P. R. China.
}

\date

\begin{abstract}
Using conventional constituent-quark model, $I=1/2$ scalar $\kappa$,
vector $K^\ast(892)$, and axial vector $K_1$ mesons are studied
in the asqtad-improved staggered fermion  with the wall-source and point-sink interpolators.
The mass ratio of $m_{\kappa}/m_{K^\ast(892)}$ is numerically confirmed
to vary apparently with quark mass,
and the experimental ordering $m_{K^\ast(892)} > m_{\kappa}$ is elegantly hold
when the light $u/d$ quark masses are sufficiently small,
while the valence strange quarks are fixed to its physical values.
We also get reasonable signals for $K_1$ meson suggested by SCALAR Collaboration from lattice QCD.
The computations are conducted with the MILC $N_f=3$ flavor gauge configurations
at three lattice spacings: $a\approx 0.15$, $0.12$, and $0.09$~fm.
\end{abstract}

\begin{keyword}
scalar meson, vector meson, axial vector meson.
\end{keyword}

\begin{pacs}
{12.38.Gc,11.15.Ha}
\end{pacs}

\begin{multicols}{2}
\maketitle

\section{Introduction}
In modern hadron physics, chiral symmetry plays an important role,
and $\sigma$ meson is a meaningful ingredient.
$\kappa$ meson can be naturally regarded as a member of
nonet scalar states of the chiral $SU(3) \bigotimes SU(3)$ symmetry 
due to the existence of $\sigma$ meson.
At present, $\sigma$ and $\kappa$ mesons are both listed
in the Particle Data Group (PDG)~\cite{Agashe:2014kda}.
Moreover, the presence of $\kappa$ meson is strongly indicated
by the recent experiments~\cite{Bugg:2009uk,Aitala:2002kr,Alde:1997ri,Bonvicini:2008jw}.

$K^\ast(892)$ meson is a well-established $q\bar q$ state belonging to $SU(3)$ octet,
some experimental analyses have precisely measured its mass~\cite{Ablikim:2005ni}.
Moreover, there are a few lattice studies on $K^\ast(892)$ meson~\cite{Aoki:2002fd,Fu:2012tj,Prelovsek:2013ela,Bernard:2001av}.
Using quenched approximation, SCALAR group
have explored $\kappa$ meson ({\it i.e.}, $I=\tfrac{1}{2}$ scalar $\bar{q}q$)
with relatively large quark mass, and found that the mass ratio of
$\kappa$ mass to $K^\ast(892)$ mass ({\it i.e.}, $m_\kappa/m_{K^\ast}$)
is about $2.0$ with rather large statistical errors~\cite{Kunihiro:2003yj,Wada:2007cp,Kunihiro:2009zz,Nakamura:2004qd},
which is not consistent with experimental ordering $m_\kappa < m_{K^\ast(892)}$.

Note that the point-source and point-sink interpolators of $I=\tfrac{1}{2}$ low-lying mesons
unavoidably leads to large statistical errors
due to the mixture of excited states,
moreover, the lattice cutoff was not properly selected
to house the large meson masses for rather small lattice size.
Consequently, the SCALAR-determined $\kappa$ masses
should be only regarded as the upper limits,
as commented by themselves in Refs.~\cite{Kunihiro:2003yj,Wada:2007cp,Kunihiro:2009zz,Nakamura:2004qd}.

Using dynamical simulations and small quark mass, SCALAR group recently
presented the preliminary result
based on the variational method
for the mass ratio of $\kappa$ mass to $\rho(770)$ mass at a single lattice ensemble,
{\it i.e.,} $m_\kappa/m_{\rho} = 1.29(5)$ \cite{Sekiguchi:2011bza},
which indicates that the mass ratio of $m_\kappa/m_{K^\ast}$
should be smaller than $1.29$.
All the SCALAR results indicate that the mass ratio $m_\kappa/m_{K^\ast}$
is definitely not constant, and rests on the quark mass.
This is actually in agreement with Nebreda and Pelaez's brilliant expositions
that $\kappa$ mass has a much stronger dependence of quark mass than that
of $K^\ast(892)$, in fact, it grows a factor of $3$ faster~\cite{Nebreda:2010wv}.

In our previous works, we follow SCALAR group's work to exploit
the point-source and point-sink interpolators to study
$K^\ast(892)$ decay width~\cite{Fu:2012tj} and $\kappa$ meson~\cite{Fu:2011zz}.
Nonetheless, as expected, we found that the statistical errors
are not quite satisfactory, and the plateaus in the mass plots are not quite clear.

The noise-to-signal ratio in the lattice computation of a meson mass
is generally proportional to $e^{(m_M-m_\pi)t}$, where $m_M$ is the desired meson mass.
Hence, the most efficient way to improve statistics is to
use smaller quark mass and finer lattice space.
Lattice studies with staggered fermion are more economical
than those with other discretizations, which permits lattice studies
with larger lattice spatial dimension ($L$) or smaller quark mass.
For these reasons, we carry out a lattice calculations on
$N_f = 3$ flavor MILC gauge configurations in the asqtad-improved staggered sea quarks.
In this simulation, the valence strange quark is fixed to its physical mass, 
and we use a broad range of light valence quark mass,
where the largest lattice dimension is $40^3 \times 96$,
and the lowest pion mass is $240$~MeV, which allow us to further explore the chiral limit.

Additionally, in order to efficiently reduce the overlaps to
the excited states of $I=1/2$ low-lying mesons
we use the wall-source and point-sink interpolators,
which are practised by MILC Collaboration to
calculate hadron spectrum~\cite{Aubin:2004wf,Bernard:2001av}.
As anticipated, the statistical errors are significantly reduced
as compared with our previous works~\cite{Fu:2012tj,Fu:2011zz}.
Moreover, we found that mass ratio of $m_\kappa/m_{K^\ast}$ is quantitatively
dependent on quark mass, and it is indeed larger than one for the larger quark mass.
The by-product of this work is that $K^\ast(892)$ mass is turned out to be larger than $\kappa$ mass
when the light $u$ quark mass is sufficiently small, thus,
the experimental ordering $m_{K^\ast(892)}>m_{\kappa}$ is elegantly hold
even we study them with simple constituent-quark model.

The high-statistics WA3 experiment established the existence of  $K_1$ meson~\cite{Daum:1981hb}.
The axial vector $K_1$ meson are now listed  in PDG
with $I(J^P)=\tfrac{1}{2}(1^+)$~\cite{Agashe:2014kda},
and the pioneering lattice studies of $K_1$ meson
were investigated by SCALAR group~\cite{Kunihiro:2003yj,Wada:2007cp,Kunihiro:2009zz,Nakamura:2004qd}.
For completeness of the study on $I=\tfrac{1}{2}$ low-lying mesons,
we also attempt at $K_1$ meson,
which is the possible candidate of
the oscillating parity partner of $K^\ast(892)$ meson
with staggered fermions~\cite{Fu:2012tj}.
In this work, the signals of $K_1$ meson are found to be acceptable.

\section{Correlator}
\label{Sec:Methods}
As explained early, to efficiently reduce the overlaps of excited states
we use the wall-source and point-sink interpolators for the $I=1/2$ low-lying mesons~\cite{Aubin:2004wf,Bernard:2001av}.
In the present study,  we use the staggered fermions,
and the local lattice operators for the vector $K^\ast(892)$ meson,
scalar $\kappa$ meson,
and axial vector $K_1$ meson are
$\gamma_i \otimes \gamma_i$, $I \otimes I$, and
$\gamma_i \gamma_5 \otimes \gamma_i \gamma_5$, respectively~\cite{Kunihiro:2005vw,Kunihiro:2007zza,vanBeveren:2006sn,Golterman:1985dz}.

It is important to note that the $\kappa$, $K^\ast$ and $K_1$ propagators
only involve a connected diagram and contain no disconnected part,
which is difficult to calculate~\cite{Kunihiro:2003yj,Bernard:2007qf}.
Moreover, $\kappa$ meson is a flavor non-singlet state,
which can not mix with the glueball state.
As a result, the signals of $\kappa$ propagators should be relatively clear,
and the lattice studies on the $I=1/2$ low-lying mesons will be important way to
shed some lights on the mystical structures of the scalar meson.

The staggered fermion meson propagators, in general, contains the single-particle form~\cite{Golterman:1985dz},
\begin{eqnarray}
{\cal C}(t) &=&
\sum_i A_i \cosh\left[m_i \left(t-\frac{T}{2}\right)\right] \cr
&+& \sum_i A_i^{\prime}(-1)^{t+1} \cosh\left[m_i^\prime \left(t-\frac{T}{2}\right)\right] ,
\label{eq:fitmodel}
\end{eqnarray}
where desired- and opposite-parity states are included,
and the opposite-parity states maintain
the temporally oscillating prefactor $(-1)^{t+1}$~\cite{Golterman:1985dz}.

For $K^\ast(892)$ correlator, only one mass is taken with
each parity~\cite{Bernard:2001av,Aubin:2004wf},
and the oscillating parity partner is a meson with $J^P=1^+$.
$K_1$ meson is also with $J^P=1^+$,
thus it is the possible parity partner of $K^\ast(892)$ meson.
Nevertheless, the states with $J^P=1^+$
can be multihadron states as well~\cite{Fu:2012tj}.
As MILC fitting $a_1$ meson,
for axial vector $K_1$ correlator,
we can, in principle, use three-state form
to get acceptable fitting quality~\cite{Bernard:2001av}.
However, in practice, we can also obtain satisfactory
fitting quality with only two-state form.
In fact, this is partially physical,
since the splittings between the ground and excited states are generally
larger for $K_1$ meson.
Furthermore, the parity partner of $\kappa$ meson is $K_A$ meson~\cite{Golterman:1985dz}.
As practised in our previous work~\cite{Fu:2011zz},
$\kappa$ correlators should be fit with the consideration of bubble comtribution.

We should remark at this point that our lattice investigation
for $K_1$ meson is absolutely an ideal case~\cite{Kunihiro:2009zz},
since $K_1$ meson is actually from the mixing between $I=1/2$ $J^{PC}=1^{++}$
and $1^{+-}$ state~\cite{Kunihiro:2009zz}, which are of
pseudoscalar-vector character~\cite{Geng:2006yb}. 

\section{Simulations and results}
\label{Sec:Results}
In the present study, we use the MILC $N_f=3$ flavor gauge configurations
with asqtad-improved staggered sea quarks~\cite{Bernard:2001av,Bazavov:2009bb,Bernard:2010fr}.
The simulation parameters of the MILC gauge configurations
are listed in Table~\ref{tab:MILC_configs},
where a broad range of quark mass is included to investigate
the mass ratios of $m_\kappa/m_{K^\ast}$, especially at small quark masses,
which are not explored by SCALAR collaboration yet.
The gauge configurations were gauge-fixed to Coulomb gauge
before calculating the propagators.
For simple notation, it is convenient to use $(am_l,am_s)$
to identify the MILC lattice ensembles.
Moreover, it is MILC's convention to call lattice ensembles as ``fine''
for spatial lattice spacing $a\approx0.09$~fm,
``coarse'' for $a\approx0.12$~fm, and ``medium-coarse'' for $a\approx0.15$~fm,
respectively.

The standard conjugate gradient method is utilized to acquire
the required matrix elements of inverse Dirac fermion matrices.
All the numerical calculations are evaluated in double precision
to avoid the potential roundoff errors,
and the conjugate gradient residual is selected to be $1.0\times10^{-5}$,
which is generally smaller than that of generating the MILC gauge
configurations~\cite{Bernard:2001av}.
Moreover, in order to improve the statistics,
all the propagators are calculated from a given number of time slices ($N_{\mathrm{slice}}$)
which are indicates in the ninth Column of Table~\ref{tab:MILC_configs}.
The time slices are evenly spread through the lattice,
namely, only one source time slice was chosen at a time.
At the end of the evaluation, we gather all the propagators.
It should be worthwhile to stress that we can adjust the values of $N_{\mathrm{slice}} $
for each lattice ensemble to
guarantee the extraction of relevant masses with desired precision.
\end{multicols}

\begin{center}
\tabcaption{
\label{tab:MILC_configs}
Lattice dimensions are described with spatial ($L$) and temporal ($T$) size.
The gauge coupling is given by $\beta = 10/g^2$.
The bare masses of the light and strange quark masses
are written in terms of $am_l^\prime$ and $am_s^\prime$, respectively.
$L=aN_L$ is the lattice spatial dimension in fm,
and pion masses are  given in MeV.
The physical strange-quark mass is indicated by $a m_s$~\cite{Bernard:2010fr}.
The number of gauge configurations used in this work is given by $N_{\mathrm{cfg}}$,
and the number of time slices calculated the propagators for each configuration
is shown by $N_{\mathrm{slice}}$.
Last Column gives our lattice-measured $\rho$ mass in lattice units.
}
\footnotesize
\begin{tabular*}{175mm}{@{\extracolsep{\fill}}lllccccccl}
\toprule
$L^3 \times T$ &$\beta$ & $am_l^\prime/am_s^\prime$ & $L{\rm[fm]}$ & $m_\pi$~(MeV)
& $am_s^\prime$ & $r_1/a$  & $N_{\mathrm{cfg}}$ & $N_{\mathrm{slice}}$  & $a m_\rho$ \\
\hline
\multicolumn {10}{c}{$a \approx 0.09$~fm}        \\
$40^3\times96$ & $7.08$   & $0.0031/0.031$  & $3.3$ & $246$
               & $0.0252$ & $3.695(4)$      & $100$ & $48$   & $0.3571(39)$ \\
$32^3\times96$ & $7.085$  & $0.00465/0.031$ & $2.8$ & $301$
               & $0.0252$ & $3.697(3)$      & $200$ & $48$   & $0.3721(43)$ \\
$28^3\times96$ & $7.09$   & $0.0062/0.031$  & $2.4$ & $347$
               & $0.0252$ & $3.699(3)$      & $200$ & $48$   & $0.3892(40)$ \\
$28^3\times96$ & $7.10$   & $0.0093/0.031$  & $2.4$ & $423$
               & $0.0252$ & $3.705(3)$      & $500$ & $16$   & $0.4051(21)$ \\
$28^3\times96$ & $7.11$   & $0.0124/0.031$  & $2.4$ & $487$
               & $0.0252$ & $3.712(4)$      & $480$ & $16$   & $0.4133(20)$ \\
$28^3\times96$ & $7.18$   & $0.031/0.031$   & $2.4$ & $756$
               & $0.0252$ & $3.822(10)$     & $484$ & $16$   & $0.4742(32)$ \\
\multicolumn {10}{c}{$a \approx 0.12$~fm}        \\
$32^3\times64$ & $6.715$  & $0.005/0.005$ & $3.7$  & $275$
               & $0.0344$ & $2.697(5)$    & $696$  & $32$  & $0.4984(24)$ \\
$24^3\times64$ & $6.76$   & $0.005/0.05$  & $2.9$  & $268$
               & $0.0344$ & $2.647(3)$    & $517$  & $32$  & $0.5309(35)$ \\
$20^3\times64$ & $6.76$   & $0.007/0.05$  & $2.4$  & $316$
               & $0.0344$ & $2.635(3)$    & $251$  & $32$  & $0.5560(30)$ \\
$20^3\times64$ & $6.76$   & $0.01/0.05$   & $2.4$  & $372$
               & $0.0344$ & $2.618(3)$    & $210$  & $4$  & $0.5724(74)$ \\
$20^3\times64$ & $6.79$   & $0.02/0.05$   & $2.4$  & $523$
               & $0.0344$ & $2.644(2)$    & $264$  & $4$  & $0.6617(28)$ \\
$20^3\times64$ & $6.81$   & $0.03/0.05$   & $2.4$  & $638$
               & $0.0344$ & $2.650(4)$    & $564$  & $4$  & $0.6483(20)$ \\
$20^3\times64$ & $6.83$   & $0.04/0.05$   & $2.4$  & $733$
               & $0.0344$ & $2.664(5)$    & $350$  & $4$  & $0.6848(30)$ \\
$20^3\times64$ & $6.85$   & $0.05/0.05$   & $2.4$  & $818$
               & $0.0344$ & $2.686(8)$    & $424$  & $4$  & $0.7154(16)$ \\
$20^3\times64$ & $6.96$   & $0.1/0.1$     & $2.4$  & $1155$
               & $0.0344$ & $2.687(0)$    & $340$  & $4$  & $0.8621(9)$ \\
\multicolumn {10}{c}{$a \approx 0.15$~fm}        \\
$20^3\times48$ & $6.566$  & $0.00484/0.0484$ & $2.9$ & $240$
               & $0.0426$ & $2.162(5)$       & $604$ & $24$  & $0.6695(78)$ \\
$16^3\times48$ & $6.572$  & $0.0097/0.0484$  & $2.3$ & $334$
               & $0.0426$ & $2.152(5)$       & $631$ & $24$  & $0.6968(59)$ \\
$16^3\times48$ & $6.586$  & $0.0194/0.0484$  & $2.3$ & $463$
               & $0.0426$ & $2.138(4)$       & $621$ & $24$  & $0.7542(53)$ \\
$16^3\times48$ & $6.600$  & $0.0290/0.0484$  & $2.3$ & $559$
               & $0.0426$ & $2.129(5)$       & $576$ & $24$  & $0.7884(37)$ \\
$16^3\times48$ & $6.628$  & $0.0484/0.0484$  & $2.3$ & $716$
               & $0.0426$ & $2.124(6)$       & $618$ & $24$  & $0.8659(18)$ \\
\bottomrule
\end{tabular*}
\end{center}
\begin{multicols}{2}

The valence $u/d$  quark masses are set to its dynamical quark
masses for all lattice ensembles,
while valence strange quark is fixed to its physical mass,
which was determined
by MILC Collaboration~\cite{Bernard:2010fr}.
In the usual manner, we extracted $\pi$, $K$,
and fictitious $s\bar{s}$ masses~\cite{Bernard:2001av}.
These pseudoscalar masses are used to evaluate
the bubble contribution to $\kappa$ correlators~\cite{Fu:2011zz},
where  three low energy couplings ($\mu$, $\delta_A$, and $\delta_V$)
are fixed to the MILC-determinated values~\cite{Aubin:2004fs}.
After neatly removing the unwanted bubble terms from $\kappa$ propagators,
the remaining $\kappa$ propagators have a clean information,
we then fits them with the physical model in Eq.~(\ref{eq:fitmodel}).

As practised in Ref.~\cite{Bernard:2001av},
the propagators of $I=1/2$ low-lying mesons
are commonly fit by changing the minimum fitting distances $\rm D_{min}$
and putting the maximum distances $\rm D_{max}$
either at $T/2$ or where the fractional statistical errors
for two consecutive time slices
are roughly beyond $30\%$.
The mean value and statistical error at each time slice are
computed by the jackknife technique.
The masses of $\kappa$, $K^\ast(892)$ and $K_1$ mesons are
secured from the effective mass plots, and they were
cautiously picked up by the overall assessment of the
plateau in the mass as the function of $\rm D_{min}$,
good fit quality, and $\rm D_{min}$ large enough to
suppress the excited states~\cite{Bernard:2001av,Fu:2011zz}.

For MILC fine, coarse, and medium-coarse lattice ensembles,
we give an example effective mass plots of
$\kappa$, $K^\ast(892)$ and $K_1$ mesons,
which are exhibited in Fig.~\ref{fig:plateau_kappa}.
We found that the plateaus for $K^\ast(892)$ meson are clear,
and the effective mass plots commonly
have small uncertainties within a broad minimum
time distance region for the fine $(0.0031,0.031)$ ensemble,
coarse $(0.005,0.05)$ ensemble,
and medium-coarse $(0.00484,0.0484)$ ensemble.
Note that  the statistics of
$K^\ast(892)$ meson is significantly improved as compared with our previous
study for the $(0.00484,0.0484)$ ensemble~\cite{Fu:2012tj},
where the point-source and point-sink interpolators are used.

From Fig.~\ref{fig:plateau_kappa}, we notice that the plateaus
for $\kappa$ meson are obvious, and are remarkably enhanced
as compared with our previous studies~\cite{Fu:2011zz},
where the point-source interpolator is used,
and its plateaus are often pretty short~\cite{Fu:2011zz}.
It is interesting and important to note that the plateaus
of $\kappa$ meson are clearly below those of $K^\ast(892)$ meson
for all three lattice ensembles,
and the effective mass plots commonly has small errors
within a relatively broad minimum
time distance region: $9<{\mathrm{D_{min}}} < 18$ for the $(0.0031,0.031)$ ensemble,
$4<{\mathrm {D_{min}}} < 15$ for the $(0.005,0.05)$ ensemble, and
$4<{\mathrm {D_{min}}} < 9$ for the $(0.00484,0.0484)$ ensemble.

\begin{center}
\includegraphics[width=8.0cm]{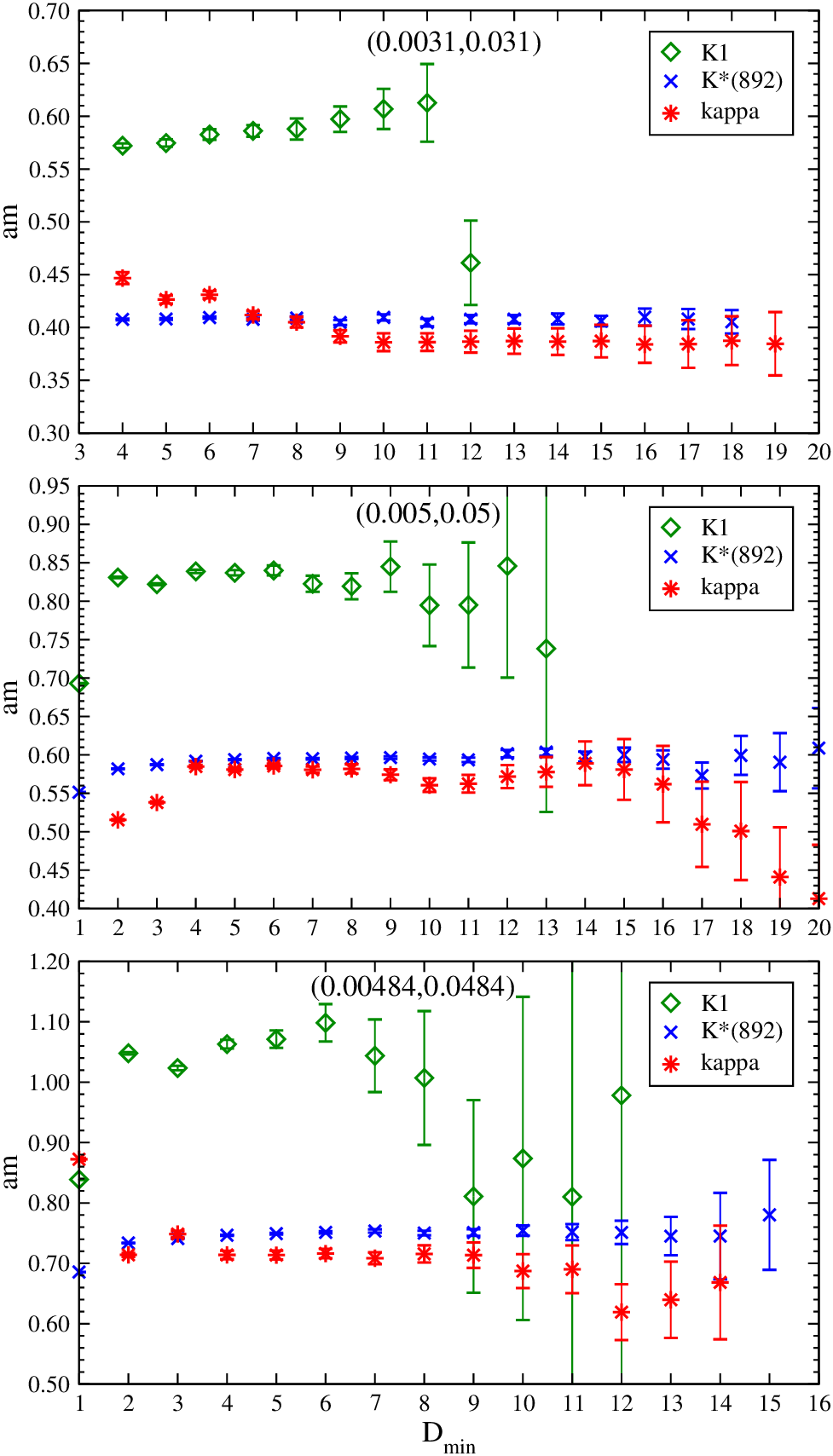}
\figcaption{
\label{fig:plateau_kappa}
Mass plots of $\kappa$, $K^\ast(892)$, and $K_1$ mesons
as a function of $D_{\rm min}$ for  MILC fine $(0.0031,0.031)$ ensemble (top panel),
the coarse  $(0.005,0.05)$ ensemble (middle panel), and
the medium-coarse $(0.00484,0.0484)$ ensemble (bottom panel).
}
\end{center}
Fitting $K_1$ meson is challenging due to its large mass.
From Fig.~\ref{fig:plateau_kappa},
we note that the plateaus in the effective mass plots are often short.
This is not surprise for us since $K_1$ meson is a $p$-wave meson~\cite{Bernard:2001av}.
In practice, we have to select the enough small $\rm D_{min}$
to get the acceptable fits,
and only the time range $4\le D_{\rm min} \le 6$ is considered.

The extracted masses of $\kappa$, $K^\ast(892)$ and $K_1$ mesons,
along with fitting ranges, fitting qualities,
are summarized in Table~\ref{tab:fitted_masses}.
The last block gives the mass ratios $m_\kappa/m_{K^\ast}$,
where the errors are inherited from the statistical errors
of both $\kappa$ and ${K^\ast(892)}$ mesons.
It is worth mentioning that that $\kappa$ mass and $K^\ast$ mass
have only small statistical errors, and the errors of $K_1$ mass are also reasonable.
This reveal that the usage of wall-source and point-sink interpolators
is a key technique in this work.
Of course, the summations over an enough number of time slices
for propagators also pay a important role.
In order to intuitively see the pion mass dependence of $\kappa$, $K^\ast(892)$ and $K_1$ mesons,
we plot these masses as the function of $(am_\pi)^2$ in Fig.~\ref{fig:pion_dependence}.
\begin{center}
\includegraphics[width=8cm]{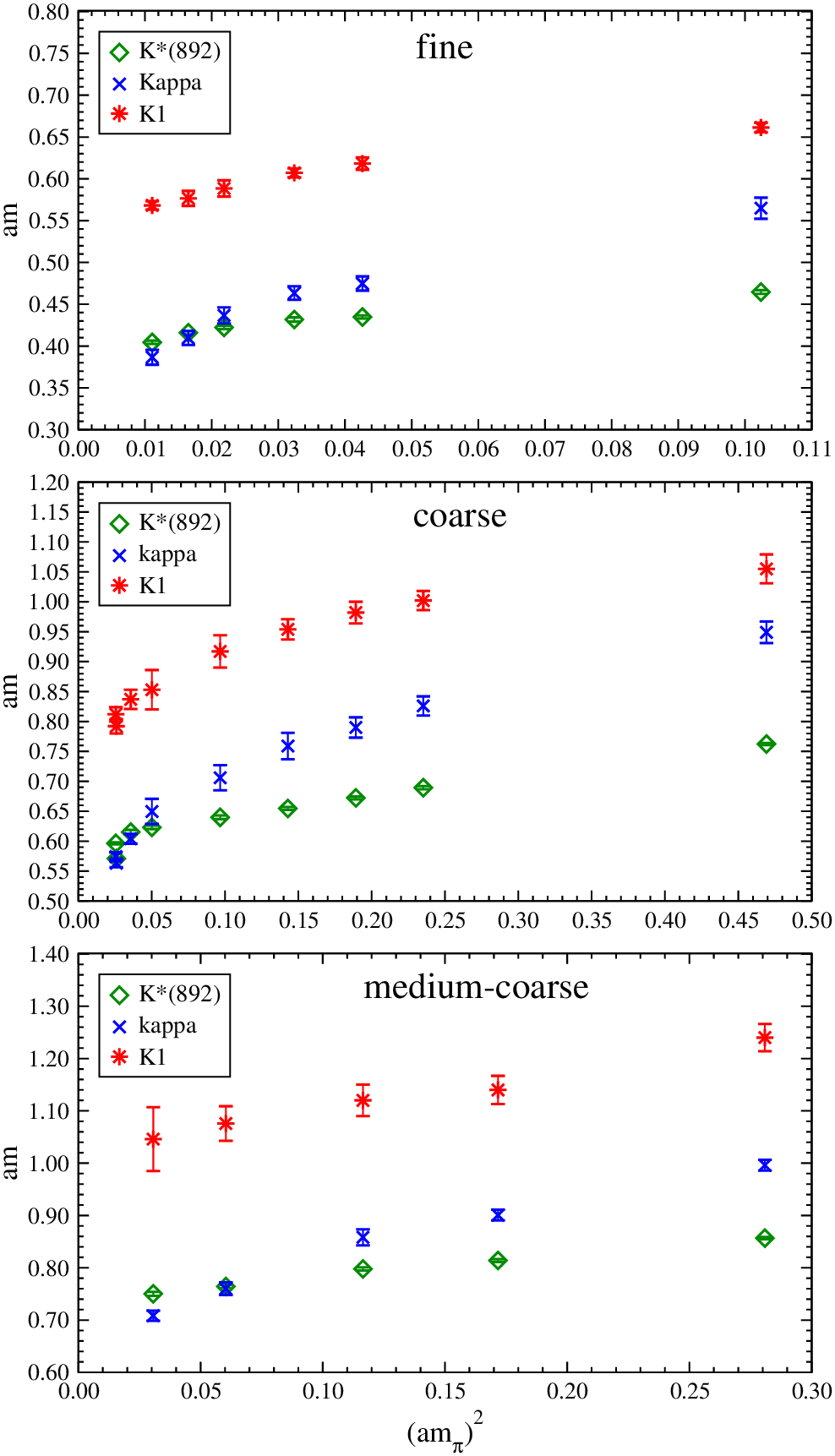}
\figcaption{
\label{fig:pion_dependence}
The masses of $\kappa$, $K^\ast(892)$, and $K_1$ mesons
as the function of $(am_\pi)^2$ for the MILC fine (top panel), coarse (middle pannel)
and medium-coarse (bottom panel) lattice ensembles.
}
\end{center}

We should remark at this point that we can not directly compare our results
for $K^\ast(892)$  meson with those of MILC determinations~\cite{Aubin:2004wf,Bernard:2001av}
since the valence strange quark masses are usually set to be equal to its sea quark by
MILC collaboration~\cite{Aubin:2004wf,Bernard:2001av},
while we fix the valence strange quark to be its physical mass in the present study.
To fairly compare our calculation, we list our lattice-measured $\rho$ meson
in last column of Table~\ref{tab:MILC_configs},
which are in well agreement with the MILC
determinations~\cite{Aubin:2004wf,Bernard:2001av}.
Moreover, our lattice-measured pion masses listed in Column $5$ of Table~\ref{tab:MILC_configs}
are also perfectly consist with the MILC
determinations~\cite{Aubin:2004wf,Bernard:2001av}.

\end{multicols}

\begin{center}
\tabcaption{
\label{tab:fitted_masses}
Summaries of $I=1/2$ low-lying meson masses. The second, third, fourth blocks
show the fitted value of $m_{K^\ast}$, its fit range,
and  fit quality: $\chi^2$ divided by number of degrees of freedom, respectively.
The fifth, sixth, and seventh blocks
give the fitted value of $m_{\kappa}$, its fit range,
and  fit quality, respectively.
The eighth, ninth, tenth blocks
give the fitted value of $m_{K_1}$, its fit range,
and  fit quality, respectively.
The last block gives the mass ratios of $m_\kappa/m_{K^\ast}$.
}
\footnotesize
\begin{tabular*}{175mm}{@{\extracolsep{\fill}}lcccccccccr}
\toprule
{\rm Ensemble}  & $a m_{K^\ast}$ & range  & $\chi^2/{\rm D}$ &
$a m_{\kappa}$ & range  & $\chi^2/{\rm D}$ &
$a m_{K_1}   $ & range  & $\chi^2/{\rm D}$ & $m_\kappa/m_{K^\ast}$ \\
\hline
$(0.0031,0.031)$ & $0.4045(19)$ & $14-48$ & $42.2/31$ & $0.3866(89)$ & $12-32$ & $19.3/17$
                 & $0.5682(99)$ & $8-16$  & $3.6/5$   & $0.956(22)$  \\
$(0.00465,0.031)$& $0.4160(20)$ & $15-48$ & $27.5/30$ & $0.4097(84)$  & $12-40$ & $36.4/25$
                 & $0.5766(90)$ & $7-24$  & $16.1/14$ & $1.006(21)$ \\
$(0.0062,0.031)$ & $0.4223(19)$ & $15-48$ & $37.1/30$& $0.4367(96)$ & $12-48$ & $23.1/33$
                 & $0.5885(97)$ & $6-16$  & $15.3/7$  & $1.034(23)$ \\
$(0.0093,0.031)$ & $0.4318(25)$ & $17-30$ & $14.6/10$ & $0.4634(81)$ & $12-48$ & $33.2/33$
                 & $0.6071(54)$ & $6-16$  & $11.6/7$  & $1.073(11)$ \\
$(0.0124,0.031)$ & $0.4348(18)$ & $15-48$ & $23.7/30$ & $0.4748(85)$ & $12-48$ & $34.7/33$
                 & $0.6184(69)$ & $6-16$  & $9.3/7$   & $1.092(20)$ \\
$(0.031,0.031)$ & $0.4646(22)$  & $17-48$ & $22.5/28$ & $0.5649(126)$ & $12-48$ & $42.6/33$
                & $0.6614(56)$  & $6-16$  & $4.1/7$   & $1.216(28)$ \\
\hline
$(0.005,0.005)$ & $0.5708(23)$ & $10-32$ & $17.5/19$ & $0.5630(65)$ & $9-32$ & $28.6/20$
                & $0.792(12)$  & $7-32$  & $26.7/22$ & $0.986(12)$ \\
$(0.005,0.05)$  & $0.5965(16)$ & $9-32$  & $26.4/20$ & $0.5742(71)$ & $9-32$ & $20.7/20$
                & $0.812(12)$  & $7-32$  & $9.8/22$  & $0.963(12)$ \\
$(0.007,0.05)$  & $0.6152(36)$ & $10-32$ & $10.3/19$ & $0.6041(82)$ & $9-32$ & $25.3/20$
                & $0.837(16)$  & $6-16$  & $5.4/7$   & $0.982(15)$ \\
$(0.01,0.05)$   & $0.6231(29)$ & $10-18$ & $3.8/5$   & $0.650(21)$ & $7-32$ & $29.9/22$
                & $0.853(33)$  & $6-16$  & $4.2/7$   & $1.043(33)$\\
$(0.02,0.05)$   & $0.6399(30)$ & $8-32$  & $18.3/21$ & $0.706(21)$  & $8-16$ & $6.8/5$
                & $0.917(27)$  & $6-16$  & $5.7/7$   & $1.110(44)$\\
$(0.03,0.05)$   & $0.6547(25)$ & $9-32$ & $22.6/20$ & $0.759(22)$  & $8-18$ & $6.3/7$
                & $0.954(17)$  & $6-20$ & $6.5/11$  & $1.163(33)$\\
$(0.04,0.05)$   & $0.6723(24)$ & $9-32$ & $22.4/20$ & $0.790(17)$  & $7-32$ & $28.0/22$
                & $0.982(18)$  & $6-18$ & $7.9/9$   & $1.174(23)$\\
$(0.05,0.05)$   & $0.6893(26)$ & $10-32$& $4.5/19$  & $0.826(16)$  & $7-32$ & $25.7/22$
                & $1.002(16)$  & $6-11$ & $0.6/2$   & $1.199(23)$\\
$(0.1,0.1)$     & $0.7623(17)$ & $10-32$& $18.8/19$ & $0.949(18)$  & $7-13$ & $6.0/3$
                & $1.055(24)$  & $7-20$ & $12.2/10$ & $1.225(23)$\\
\hline
$(0.00484,0.0484)$ & $0.7502(38)$ & $8-24$  & $3.9/13$  & $0.7083(97)$   & $7-24$ & $14.2/14$
                   & $1.046(61)$  & $7-16$  & $3.6/6$   & $0.944(14)$ \\
$(0.0097,0.0484)$  & $0.7642(30)$ & $7-24$  & $17.5/14$ & $0.7601(121)$  & $7-24$ & $17.1/14$
                   & $1.076(33)$  & $6-16$  & $9.6/7$   & $0.995(16)$ \\
$(0.0194,0.0484)$  & $0.7979(30)$ & $8-24$  & $18.6/13$ & $0.8582(153)$  & $7-24$ & $11.4/14$
                   & $1.120(30)$  & $6-16$  & $3.2/7$   & $1.076(20)$ \\
$(0.0290,0.0484)$  & $0.8138(28)$ & $8-24$  & $24.6/13$ & $0.9007(102)$  & $6-24$ & $16.8/15$
                   & $1.140(27)$  & $6-16$  & $5.0/7$   & $1.107(13)$ \\
$(0.0484,0.0484)$  & $0.8567(20)$ & $8-24$  & $17.0/13$ & $0.9961(104)$  & $6-24$ & $21.7/15$
                   & $1.240(26)$  & $6-16$  & $11.5/7$  & $1.163(12)$ \\
\bottomrule
\end{tabular*}
\end{center}

\begin{multicols}{2}

From Table~\ref{tab:fitted_masses} and Fig.~\ref{fig:pion_dependence},
we clearly found that mass ratios of $m_{\kappa}/m_{K^\ast}$ definitely
vary with quark  mass, especially at lower quark masses. 
Note that it is not a crude linear form on the whole,
but on large quark masses, we can roughly approximate it in a linear form.
On small quark masses, the chiral logarithms are keenly discerned. To make our reports more intuitive, these mass ratios
are also presented graphically in Fig.~\ref{fig:massRatioR1} as a function of pion mass.
Note that the mass ratios of $m_{\kappa}/m_{K^\ast}$ are dimensionless quantities,
and we calculate these ratios at three MILC lattice spacings,
consequently it is appropriate to exhibit our results
in terms of dimensionless quantities,
and pion masses are then scaled with the MILC scale $r_1$~\cite{Bernard:2001av}. 
The definition of $r_1$ and the benefits of using $r_1$ is  discussed in Ref.~\cite{Bernard:2001av}.

As expected, we indeed reproduce the relatively large mass ratios of
$m_{\kappa}/m_{K^\ast}$ for large quark masses,
which is fairly consistent with the SCALAR group's recent tentative results,
where the mass ratio of $m_{\kappa}/m_{K^\ast}$ are indicated to be smaller than $1.29(5)$
for a single lattice ensemble with the hopping parameter $h_{u/d}=0.1390$~\cite{Sekiguchi:2011bza}.
Note that the mass ratio of $m_{\kappa}/m_{K^\ast}$ for the SCALAR group's early lattice
results~\cite{Kunihiro:2003yj,Wada:2007cp,Kunihiro:2009zz},
have a rather large statistical error (about $30\%$),
this is partially due to the usages of large quark masses,
and the point-source and point-sink interpolators.
Our previous works about $\kappa$ meson also suffered from the huge statistical error
due to the usage of the point-source and point-sink interpolators~\cite{Fu:2012tj,Fu:2011zz}.

\begin{center}
\includegraphics[width=8.5cm,clip]{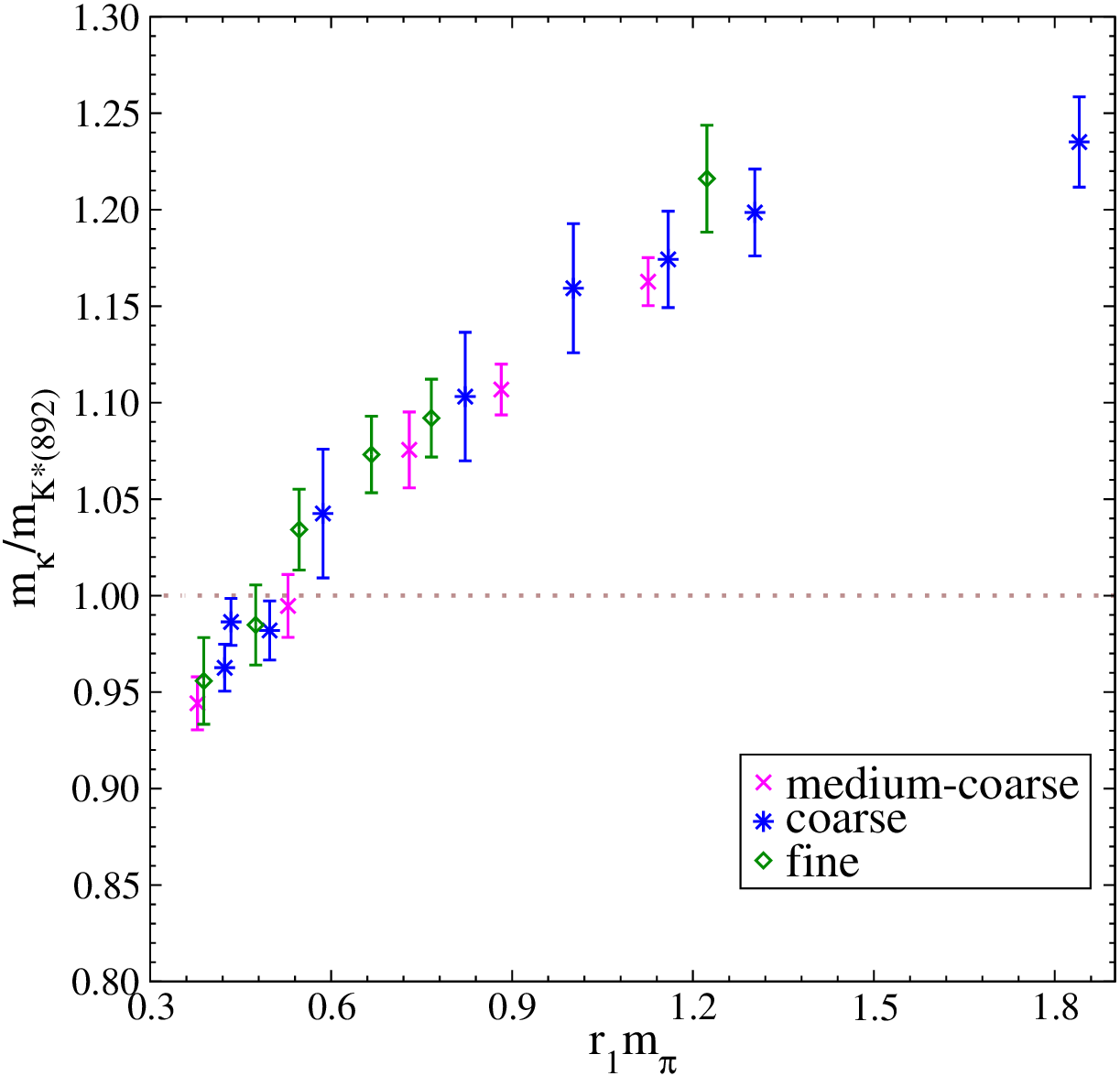}
\figcaption{
\label{fig:massRatioR1}
Mass ratios of $m_{\kappa}/m_{K^\ast}$
as a function of $r_1 m_\pi$ with the MILC scale $r_1$~\cite{Bernard:2001av}
for MILC ``fine'', ``coarse'' and ``medium-coarse'' lattice ensembles.
The brown dotted line indicates the position of $m_{\kappa}=m_{K^\ast}$.
}
\end{center}

It is interesting and important to note that mass ratios of $m_{\kappa}/m_{K^\ast}$
are small than one when the quark masses are small enough.
This is not surprise for us at all.
It is well-known that the vector $\rho$ meson grows slower than
the scalar $\sigma$ meson with the pion mass varying~\cite{Jeltema:1999na}.
Likewise, the vector $K^\ast(892)$ should grow slower than the scalar $\kappa$ meson.
Our lattice simulation indeed illustrate that $K^\ast(892)$ meson
grows slower than $\kappa$ meson when the pion mass is varied
(the valence strange quarks are fixed to its physical ones),
which is it is qualitatively consistent with Nakamura's lattice results
in Fig.~6 of Ref.~\cite{Nakamura:2004qd}.
Moreover, these features are also qualitatively consistent with Nebreda and Pelaez's
statements that $\kappa$ mass has a much stronger dependence of the quark mass
than that of $K^\ast(892)$ meson~\cite{Nebreda:2010wv}.
Consequently, $K^\ast(892)$ mass is larger than $\kappa$ mass
when pion mass is sufficient small,\footnote{
We observe the rule of thumb that mass ratios of $m_{\kappa}/m_{K^\ast}$
are small than one when pion masses are roughly small than $300$ MeV,
where the valence strange quarks are fixed to its physical ones.
}
therefore, the experimental ordering $m_{K^\ast(892)}>m_{\kappa}$ is nicely kept
even we study them with the simple constituent-quark model.
This is an encouraging and exciting result.

\section{Summary and outlook}
\label{Sec:Conclusions}
In this work, we employ the simple constituent quark model to
study $I=1/2$ low-lying $\kappa$, $K^\ast$ and $K_1$ mesons
using the MILC $N_f=3$ flavors gauge configurations
with asqtad-improved staggered sea quarks.
We employ the wall-source and point-sink interpolators
for $\kappa$, $K^\ast$ and $K_1$ mesons,
and the signals for $\kappa$ and $K^\ast$ mesons are found to be
significantly improved as compared with these using point-source and point-sink 
interpolators~\cite{Kunihiro:2003yj,Wada:2007cp,Kunihiro:2009zz,Nakamura:2004qd,Fu:2011zz},
and also can comparable with recent results of SCALAR group with
variational method~\cite{Sekiguchi:2011bza}.

We should remember that the effective mass plots of $\kappa$ meson
still suffers from the large errors at large $\rm D_{min}$.
Moreover, the signals of $K_1$ meson are not satisfactory.
Our lattice attemptation of $K_1$ meson are absolutely in the preliminary stage.
This is not surprise for us since $K_1$ meson  
are mixture of two $SU(3)$ eigenstates,
which have pseudoscalar-vector character~\cite{Geng:2006yb}.  
Additionally, $K_1$ meson is unstable and 
usually lead to new levels from the decay products~\cite{Roca:2012rx}.
These natures should be appropriately incorporated in the $K_1$ interpolators
for more sophisticated lattice simulation.

In this paper, $\kappa$ mass and $K^\ast(892)$ mass are secured with acceptable quality,
and consequently mass ratio of $m_{\kappa}/m_{K^\ast}$ is obtained with relatively small errors.
As anticipated, the mass ratio of $m_{\kappa}/m_{K^\ast}$ is numerically
confirmed to vary with pion mass,
and the experimental ordering $m_{K^\ast(892)}>m_{\kappa}$ is nicely hold
for enough small light quark mass.
This will somewhat aid researchers to examine
the intrinsic attribute of scalar meson.

Admittedly, since the strange quark mass $m_s$ is larger than the light up quark mass $m_u$,
the observed mass ordering $m_{a_0(980)} > m_\kappa$
can not be reconciled with the conventional $\bar ud$ and $\bar us$ states.
The tetraquarks interpretations of scalar mesons can realize this experimental ordering~\cite{Alford:2000mm,Prelovsek:2010kg,Loan:2008sd,Wagner:2013nta,Alexandrou:2012rm,Guo:2013nja,Wagner:2013jda}.
It should be worthwhile to stress that
the properties of the scalar resonance $a_0(980)$, whose mass is
in the vicinity of $K\bar K$ threshold,
should consider $K\bar K$ scattering, which is suggested by
ETM Collaboration in Ref.~\cite{Alexandrou:2012rm}.
Recently, there is a lattice attempt at $K\bar K$ scattering~\cite{Fu:2012ng}.
Nonetheless, the robust calculations of $K\bar K$ scattering should adopt
the finer lattice ensemble and small light quark masses~\cite{Fu:2012ng},
thus, this kind of task needs astronomical computer resources,
we reserve this ambitious work in the future.

Additionally, from Table~\ref{tab:fitted_masses} and Fig.~\ref{fig:pion_dependence}, 
we discern that the signals of $\kappa$  and $K_1$  
for lattice ensembles with small pion mass are commonly better.
Moreover, the bubble contribution to $\kappa$ operator
should be further suppressed for enough lattice spatial extent $L$.
For these purposes, we are preparing a lattice study
with the MILC superfine lattice ensembles.
Furthermore, we are also planing to use the variational method,
which includes  the traditional quark-antiquark operators and four-quark structure,
to study the scalar mesons~\cite{Wagner:2013jda,Sekiguchi:2011bza}.
We expect the signals of the relevant propagators will be significantly further improved.

The reliable extraction of $\kappa$ and $K^\ast(892)$ masses
from two-point correlation functions will inspire us
to further investigate resonance masses
of $K^\ast(892)$ and $\kappa$~\cite{Fu:2012tj,Fu:2011xw}
since both $K^\ast(892)$ and $\kappa$ mesons are resonances.
Moreover, this will stimulate us to study $\sigma$ meson
from lattice QCD~\cite{Bernard:2007qf}.
We will unceasingly and undauntedly appeal for the computer resources
to accomplish these fascinating and prospecting enterprises.

\acknowledgments{
We deeply appreciate MILC Collaboration
for using MILC gauge configurations and MILC codes.
We thank NERSC for providing a platform
to download lattice ensembles.
We sincerely thank Carleton DeTar
for inculcating me in the necessary knowledge for this work.
We appreciate K. F. Liu and Han-qing Zheng for their stimulating discussions
about scalar meson.
We especially thank Eulogio Oset for his enlightening comments.
We cordially express our boundless gratitude
to Hou qing's strong support and Fujun Gou's vigorous support,
otherwise, it is impossible for us to conduct this ambitious work.
We thanks Ning Huang for providing some devices to store propagators.
The author would like to express his gratitude to the Institute of
Nuclear Science $\&$ Technology, Sichuan University,
where computer resources  are furnished.
Numerical calculations for this work were carried out at AMAX,
CENTOS and HP workstations.
}

\newpage
\end{multicols}
\vspace{-1mm}
\centerline{\rule{80mm}{0.1pt}}
\vspace{2mm}

\begin{multicols}{2}

\end{multicols}

\clearpage

\end{CJK*}
\end{document}